\documentclass[referee]{raa}           

\usepackage{graphicx,times}
\usepackage{natbib}
\usepackage{amssymb,amsmath}
\bibpunct{(}{)}{;}{a}{}{,}

\usepackage[pagebackref=true]{hyperref}
\newcommand{\PC}{{\sc Pencil Code}~}

\begin{document}

   \title{Data-driven simulations of magnetic field evolution in Active Region 11429: Magneto-friction method using \PC}

 \volnopage{ {\bf 20XX} Vol.\ {\bf X} No. {\bf XX}, 000--000}
   \setcounter{page}{1}

   \author{P. Vemareddy
         \inst{1}, 
         J\"orn~Warnecke
         \inst{2},
         Ph.~A.~Bourdin
         \inst{3}
   }


\institute{ Indian Institute of Astrophysics, Sarjapur road, II Block, Koramangala, Bengaluru-560 034, India {\it vemareddy@iiap.res.in}\\
        \and
             Max-Planck-Institut f\"ur Sonnensystemforschung, Justus-von-Liebig-Weg 3, D-37077 G\"ottingen, Germany\\
	\and
	  Institute of Physics, University of Graz, Universit\"atsplatz 5, A-8010 Graz, Austria \\
\vs\no
   {\small Received 2023 October 16; accepted 2023 December 18}
}

\abstract{Coronal magnetic fields evolve quasi statically over long time scales and dynamically over short time scales. As of now there exists no regular measurements of coronal magnetic fields, and therefore generating the coronal magnetic field evolution using the observations of the magnetic field at the photosphere is of fundamental requirement to understand the origin of the transient phenomena from the solar active regions. Using the magnetofriction (MF) approach, we aim to simulate the coronal field evolution in the solar active region 11429. The MF method is implemented in open source \PC along with a driver module to drive the initial field with different boundary conditions prescribed from observed vector magnetic fields at the photosphere. In order to work with vector potential and the observations, we prescribe three types of bottom boundary drivers with varying free-magnetic energy. The MF simulation reproduces the magnetic structure, which better matches to the sigmoidal morphology exhibited by AIA images at the pre-eruptive time. We found that the already sheared field further driven by the sheared magnetic field, will maintain and further build the highly sheared coronal magnetic configuration, as seen in AR 11429. Data-driven MF simulation is a viable tool to generate the coronal magnetic field evolution, capturing the formation of the twisted flux rope and its eruption. 
}

   \authorrunning{Vemareddy et al. }            %author_head in even pages
   \titlerunning{Magnetofriction method with \PC}  % title_head in odd pages
   \maketitle

%________________________________________________ sections below
% 
\section{Introduction}
\label{Intro} 
%\linenumbers
Solar eruptive events like flares and CMEs are energetic, large-scale phenomena of scientific interest due to their significant impact on space weather. From several studies based on observations and numerical modeling \citep{klimchuk2001, priest2002}, it has been established that these large-scale events are magnetically driven by the storage of magnetic energy and helicity in the coronal volume of the active regions (ARs). From this point of view, understanding the structure and evolution of the coronal magnetic field have become an important element in revealing the physical origins of these events on the Sun. 

As of now, regular measurements of the magnetic fields are available at the photospheric surface \citep{Scherrer1995_MDI,schou2012} at high resolution and cadence. Continuous vector field observations of magnetic fields from \textit{Helioseismic and Magnetic Imager} on board the Solar Dynamics Observatory \citep{hoeksema2014} had been used in several studies of the magnetic field dynamics, relating their connection to the coronal features present before the occurrence of the flares/CMEs. Since the magnetic fields are line-tied, they become twisted and sheared by the corresponding photospheric plasma motions. These signatures of stored energy magnetic configurations are reported to exist in several ARs exhibiting shearing and rotating motions of the sunspot polarities \citep{ambastha1993,brown2003, tian2006, vemareddy2012_sunspot_rot}. The configurations with stored magnetic energy are referred to as magnetic nonpotentiality, which was used to quantified in terms of parameters such as the magnetic shear \citep{Hagyard1986, WangT1994}, horizontal gradient of longitudinal magnetic field \citep{Falconer2003, SongH2006, Vasantharaju2018}, electric current \citep{WangT1994, leka1996}, average force-free twist parameter $\alpha_{av}$ \citep{pevtsov1994, hagino2004,vemareddy2012_sunspot_rot}, magnetic free energy \citep{Metcalf2005a}, etc. Further, the magnetic non-potentiality was suggested to be linked with sigmoidal-shaped plasma loops in the EUV or soft X-ray observations of the corona and are the precursor \citep{rust1994, rust1996, canfield2007}.  Using the continuous vector magnetic field observations, it was possible only recently to show that the coronal accumulation of magnetic energy and helicity is predominant in the ARs exhibiting shearing and rotating motions \citep{Liuyang2012_HelEner,Vemareddy2015_HelEne, vemareddy2019_VeryFast, DhakalS2020} and those ARs produce violent coronal activity.

Since routine observations of the coronal magnetic fields are not available, the photospheric magnetic field measurements are typically extrapolated into the corona to reconstruct the 3-dimensional (3D) magnetic structure of the AR corona \citep{Sakurai1989_MagField_ext}. The coronal field is approximated as force-free owing to low beta plasma \citep{WiegelmannSakurai2012} and its evolution is modeled as the series of quasi-static force-free fields. These extrapolated fields are employed to study the coronal magnetic structure containing the null points and flux rope topology \citep{mason2009, vemareddy2014_Quasi_Stat, Guoy2016, Vemareddy2018} and are judged to be reproduce structures resembling the observed coronal plasma loops to a high accuracy \citep{Schrijver2008_nlff,Vemareddy2018a,vemareddy2019_VeryFast}. 
However, the extrapolated magnetic fields represent static fields where the dynamic evolution is missing. Magnetohydrodynamic (MHD) simulations are a viable tool to generate the dynamical evolution of the magnetic fields in the AR corona. These simulations involve advancing the full MHD equations in time and require information about plasma flow, density and temperature in addition to the magnetic field \citep[e.g.][]{Mikic1999_mhd_globalcor,GN02,GN05b,GN05a,BP11} in the computational domain. These types of models were very successful in reproducing the magnetic field and loop structure above active regions \citep{BBP13, WP19}. The type of coronal heating in such observationally driven models seems more consistent with MHD turbulence than with the dissipation of Alfv\'en waves \citep{Bourdin2016_scaling_laws}.

Although it is physically realistic to capture various processes like flare reconnection, it is computationally expensive to perform on AR scale over a timescale of a few days, even if one enhances the time step significantly \citep{WB2020}. As the magnetic field in a coronal model is almost force-free between about 6 and 60 Mm in height \citep{PWCC15,BSB2018}, alternative methods that may be used to model the coronal field are the ambipolar-diffusion and magneto-frictional approaches. While ambipolar diffusion would include only a magnetic resistivity parallel to the field $\mathbf{B}$, this method focuses on relaxing an initially known coronal field configuration. We see that this relaxation is sufficient to generate helicity \citep{BB2018}. The magneto-frictional approach uses a different resistivity term in the induction equation that depends on the current density $\mathbf{J}$ that could still point in any direction.

The magneto-frictional method \citep{Yang1986_MagFric} can be used to simulate a continuous time series of force-free fields by evolving an initial coronal field by changing the photospheric magnetic field continuously \citep{Mackay2011_ModDis}. In this approach, only the induction equation is used with the assumption that the plasma velocity is proportional to the Lorentz force \citep{Mackay2011_ModDis} and therefore the initial field relaxes to a force-free field by advancing the magnetic induction equation. The initial field evolves with the corresponding change of the observed lower boundary and preserves the magnetic connectivity and flux from one time instance to the later smoothly. These simulations have been shown to be effective in capturing comparable coronal field resembling morphology in EUV images, which is utilised to study the coronal field evolution involving filament channel and flux rope formations \citep{Cheung2012_MF,Gibb2014_mf, Pomoell2019_EleFld, Keppens2023_AMRVAC}. Compared to full MHD, the magnetofrictional simulations are computationally less expensive and typically much faster; hence, they can be used to study the evolution of coronal fields over long time scales. 

In this study, by using the magnetofriction approach, we simulate the coronal field evolution in the AR 11429. The method is implemented in \PC \citep{2021JOSS....6.2807P}, which is open source and freely available under GitHub\footnote{\url{http://github.com/pencil-code}}. We drive the initial field with different boundary conditions prescribed from observed vector magnetic fields and then analyze the coronal field evolution over a span of 3 days. The simulation method and numerical implementation is described in Sect.~\ref{NumMethod}, the output results of the simulation are presented in Sect.~\ref{Res} and a summary and conclusions are given in Sect.~\ref{Summ}.     

\begin{figure*}
	\centering
	\includegraphics[width=.6\textwidth,clip=]{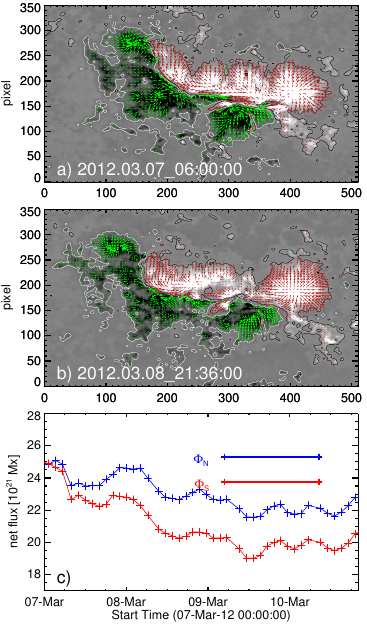}
	\caption{a-b) HMI vector magnetic field observations of the AR 11429 at two different times of the evolution. The background image is the normal component $B_z$ of the magnetic field with overplotted contours at $\pm$120 G. Arrows refer to horizontal fields, ($B_x$, $B_y$), with their length being proportional to the magnitude $B_h=\sqrt{B_x^2+B_y^2}$. The axes units are in pixels of 0.5 arcsec. c) time evolution of the net flux in each polarities of the AR. }
	\label{Fig1}
\end{figure*}

\section{Numerical Simulation }
\label{NumMethod}
\subsection{Magneto-Friction Method}
The coronal magnetic field evolution of the AR 11429 is simulated by magneto-frictional (MF) relaxation method \citep{Yang1986_MagFric}. It is a special case of the more general MHD relaxation method, which tries to solve the momentum and magnetic induction equations to obtain an equilibrium state. In the MF method, the magnetic field evolves in response to the photospheric foot point motions through the non-ideal induction equation    

\begin{equation}
	\frac{\partial \mathbf{A}}{\partial t}= \mathbf{v}_{\rm MF}\times\mathbf{B}-\eta \mu_0 \mathbf{J}
	\label{eq_Ind}
\end{equation}
where $\mathbf{v}_{\rm MF}$ is the magnetic frictional plasma velocity, $\mathbf{A}$ is the vector potential relating the magnetic field as $\mathbf{B}=\nabla\times\mathbf{A}$, and $\mathbf{J}=\nabla\times\mathbf{B}/\mu_0$ is the electric current. Solving the induction equation in terms of $\mathbf{A}$ ensures that $\nabla\cdot\mathbf{B}=0$ is always fulfilled. We choose a magnetic diffusivity of $\eta=2\times10^{8}\,$m/s$^2$ to be able to run the simulation stable. Under the assumption of magneto-static conditions, the momentum balance equation gives the plasma velocity as given by
\begin{equation}
	\mathbf{v}_{\rm MF}=\frac{1}{\nu}\frac{\mu_0\,\mathbf{J}\times\mathbf{B}}{B^2}
	\label{eq_v_mf}
\end{equation}
where $\nu$ is the magneto-frictional coefficient controlling the speed of the relaxation process and $\mu_0$ is the magnetic permeability in vacuum. As suggested in \citet{Cheung2012_MF}, the frictional coefficient $\nu$ can not be uniform because it will lead to unphysical $\mathbf{v}_{\rm MF}$ at the lower boundary, and therefore we used a height-dependant functional form given by
\begin{equation}
	\frac{1}{\nu} = \frac{1}{\nu_0}(1-e^{-z/L}),
	\label{eq_nuprof}
\end{equation}
where $\nu_0$ is set around $80\times10^{-12}$ s\,m$^{-2}$ and $z$ is the height above the bottom boundary, and $L$ is the chosen as 15 Mm. This form of frictional coefficient gives MF velocities smoothly reduces to zero towards bottom boundary $z=0$.     

%\subsection{Pencil-Code Implementation}
\subsection{\PC Implementation}
The \PC is a high-order finite-difference code for compressible hydrodynamic flows with magnetic fields. It is highly modular and MPI parallelized to run on massively supercomputers. To achieve high numerical accuracy, the code uses sixth-order in space and third-order in time differentiation schemes. More details on computational aspects are referred to \citet{Axel2003_CompAspects}. In order to ensure solenoidality, magnetic fields are implemented in terms of vector potential $\mathbf{A}$. We implement the MF relaxation technique (Eqs.~\ref{eq_v_mf} and~\ref{eq_Ind}) in the \PC. This simulation requires solving only induction equation involving the magnetic field and MF velocity; therefore, \texttt{magnetic.f90} is used, switching off all other modules corresponding to momentum equation (hydro module), continuity equation (density module), equation of state, entropy equations, etc. The code in this implementation includes calculating the pencils of MF velocity (Eq.~\ref{eq_v_mf}) using the Lorentz force and then updating the vector potential $\mathbf{A}$ with the terms in Eq.~\ref{eq_Ind}. A height-dependant friction coefficient (Eq.~\ref{eq_nuprof}) is also added in the code.

A special driver module is developed to load the bottom boundary magnetic field observations and then feed the simulation as the time progresses in steps of integration time. We use a modification of the implementation by \citep{BP11,Bourdin2020,WB2020}. The time step $\delta t$ is normally specified as Courant time step through the coefficients $c_\delta t (=0.9)$, $c_{\delta t, v} (=0.25)$ as given by      
\begin{equation}
	\delta t=min\left ( c_{\delta t} \frac{{\delta x}_{min}}{U_{max}}\,\,c_{\delta t,v} \frac{{\delta x}^2_{min}}{D_{max}} \right )
\end{equation} 
where ${\delta x}_{min}$ is the minimum grid size in all three directions, $U_{max}$ is the maximum resultant velocity and $D_{max}$ is the maximum of the diffusion coefficients used in addition to the magnetic diffusion $\eta$. To smooth the small-scale local gradients, we also include third-order hyper diffusion. % U_max included V_mf

\subsection{Initial condition}

The initial 3D magnetic field is computed with the potential field (PF, \citealt{gary1990}) as well as non-linear force-free field (NLFFF, \citealt{Wheatland2000,wiegelmann2004}) model assumption from the observed boundary condition of the AR 11429. The NLFFF is more suitable to a non-potential field of the AR with twisted structures like flux ropes or sigmoids. Using the observed boundary magnetic field of the AR at time 03:00 UT on March 7, 2012, we computed the PF and NLFFF on a uniform Cartesian computational grid of $192\times192\times120$ encompassing the AR corona of physical dimensions $ 280\times280\times175 $ Mm$^3$. Here, to reduce the computation time of AR evolution of 72 hours, we rebin the actual observations by a factor of four.  \\                       
Since the \PC runs on the vector potential instead of magnetic field, we construct the vector potential $\mathbf{A}$ of the computed 3D magnetic field $\mathbf{B}$ according to the formalism prescribed in \citet{DeVore2000}  
\begin{equation}
	\mathbf{A}(x,y,z)=\mathbf{A}_{\rm p}(x,y,0)-\hat{z}\times\int_0^z dz' \mathbf{B}(x,y,z').
\end{equation}
where $\mathbf{A}_{\rm p}(x,y,0)$ is vector potential of PF at the bottom boundary ($z=0$) and is computed from the normal or vertical component of magnetic field $B_z$ as
\begin{equation}
	\mathbf{A}_{\rm p}(\mathbf{x})=\frac{\hat{z}}{2\pi}\times \int_{S} B_z(\mathbf{x'})\frac{\mathbf{x}-\mathbf{x'}}{\left| \mathbf{x}-\mathbf{x'} \right|^2}\,dS'
\end{equation}
by imposing the conditions $\nabla\cdot\mathbf{A}_p=0$ and $\hat{n}\cdot \mathbf{A}_p=0$. These conditions are referred as coulomb gauge and DeVore gauge respectively \citep{Valori2016}.

An alternative Fourier method (e.g., \citealt{Chae2001_ObsDetMagHel}) is used to calculate the $\mathbf{A}_{\rm p}|_{z=0}$ as given by
\begin{eqnarray}
	A_{{\rm p},x}={\rm FT}^{-1}\left[ \frac{ik_y}{k_x^2+k_y^2} {\rm FT} (B_z)\right] \nonumber \\
	A_{{\rm p},y}={\rm FT}^{-1}\left[ \frac{-ik_x}{k_x^2+k_y^2} {\rm FT} (B_z)\right]
	\label{eq_ap}
\end{eqnarray}
where $k_x$, $k_y$ are wave numbers in x and y directions respectively and FT the Fourier transform.

\begin{figure*}[!ht]
\centering
\includegraphics[width=.9\textwidth,clip=]{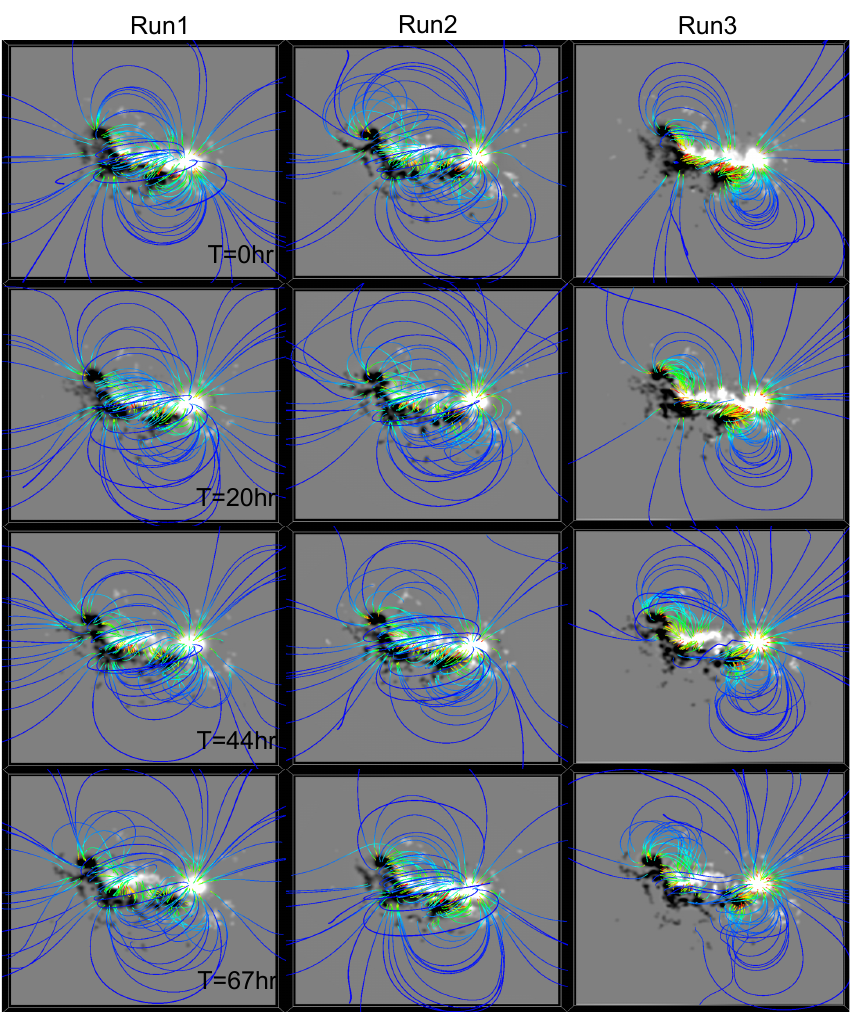}
\caption{Magnetic structure at different epochs of the simulation Run1 (first column), \texttt{Run2} (second column) and \texttt{Run3} (third column). The background image is the normal magnetic field $B_z$ at the photosphere ($z=0$) in grey shades together with magnetic field lines color-coded by their field strengths. In \texttt{Run1} and \texttt{Run2}, the magnetic field evolves to be mildly sheared as it is not much different from the initial condition. However, the magnetic structure in \texttt{Run3} is highly sheared and resembles an inverse-S shape as a whole. An animation of this figure is available online. }     
\label{Fig2}
\end{figure*}

\subsection{Boundary conditions}
\label{bc}
The time-dependent bottom boundary condition (\texttt{bc}) is prepared from the photospheric vector magnetic field observations of the AR 11429 obtained from the Helioseismic and Magnetic Imager (HMI, \citealt{schou2012, bobra2014}) onboard the Solar Dynamics Observatory. These observations have spatial resolution of 0.5"/pixel and temporal cadence of 720s. These magnetic field observations are smoothed with a Gaussian width of 3 pixels, both spatially and temporally. Flux balance condition is also imposed on each magnetogram by multiplying the flux of one polarity by the factor that it is smaller or larger than the other polarity. These observations are then spatially re-binned by a factor of four, so the resolution of the simulation roughly becomes 2"/pixel (1.465 Mm). From these observations, we prescribe three types of boundary conditions, in the form of vector potential, for three different runs.

\begin{enumerate}
	\item \texttt{bc1}: The simplest bc for the bottom boundary is normal component of magnetic field which is written as vector potential $\mathbf{A}_{\rm p} (x,y)$ (Eq.~\ref{eq_ap}). With this bc, the changing boundary as the AR evolves, embeds the foot point motion and then the coronal field evolves correspondingly in the MF relaxation. This bc was used in the most of the earlier works using the MF approach \citep{Mackay2011_ModDis, Yardley2018_MF_11437} or the full MHD simulations \citep[e.g.][]{BP11,BBP13,WP19}.
	\item \texttt{bc2}: In order to include observed twist information from the horizontal components of the magnetic field, we prepared a bc using
	\begin{equation}
		\mathbf{A}(x,y,z=1)=\mathbf{A}_{\rm p}(x,y,z=0)-\hat{z}\times\int_0^{z=1} dz' \mathbf{B}_{\rm lff}(x,y,z')
	\end{equation}
	where $\mathbf{B}_{\rm lfff}$ is the linear force-free field described by the average AR magnetic twist $\alpha_{\rm av}$. We use this \texttt{bc} at $z=0$ instead at $z=1$ layer which is at $\delta z=1.456$ Mm in height. This modification does not alter the bottom $B_{\rm z}$ more than 0.1\% and is necessary to facilitate injecting AR twist into the corona when working with vector potential instead of magnetic field directly. 
	\item \texttt{bc3}: The twist information can also be invoked into the \texttt{bc} by using the direct observations of horizontal field 
	\begin{equation}
		\mathbf{A}(x,y,z=1)=\mathbf{A}_{\rm p}(x,y,z=0)-\hat{z}\times\int_0^{z=1} dz' \mathbf{B}_{\rm obs}(x,y,z') 
	\end{equation}
	where the $\mathbf{B}_{\rm obs}|_{z=1}$ is obtained with NLFFF extrapolations from $\mathbf{B}_{\rm obs}|_{z=0}$. Since the NLFFF is also computationally expensive to perform at every time instant of the observations, one can make an approximation that 
	%    $\mathbf{B}_{\rm obs}|_{z=1} = fact*\mathbf{B}_{obs}|_{z=0}$, where the \texttt{fact} is the factor by which the magnetic field falls from $z=0$ to $z=1$ layer. In most of the extrapolation models, the magnetic fields are found to decrease exponentially from the photosphere into the corona, so we set \texttt{$fact=0.7$} in our setup of bc at $z=0$.
	$\mathbf{B}_{\rm obs}|_{z=1} = c\,\mathbf{B}_{obs}|_{z=0}$, where the $c$ is the factor by which the magnetic field falls from $z=0$ to $z=1$ layer. In most of the extrapolation models, the magnetic fields are found to decrease exponentially from the photosphere into the corona, so we set $c=0.7$ in our setup of \texttt{bc} at $z=0$.
\end{enumerate}

These boundary conditions at z=0 are extrapolated using potential field model into ghost layers below bottom boundary. Since the derivatives are 6th order accurate, three ghost layers are included to evaluate the derivatives near the boundaries. In the horizontal direction, we adopt periodic boundary conditions. 

\begin{figure*}
\centering
\includegraphics[width=.69\textwidth,clip=]{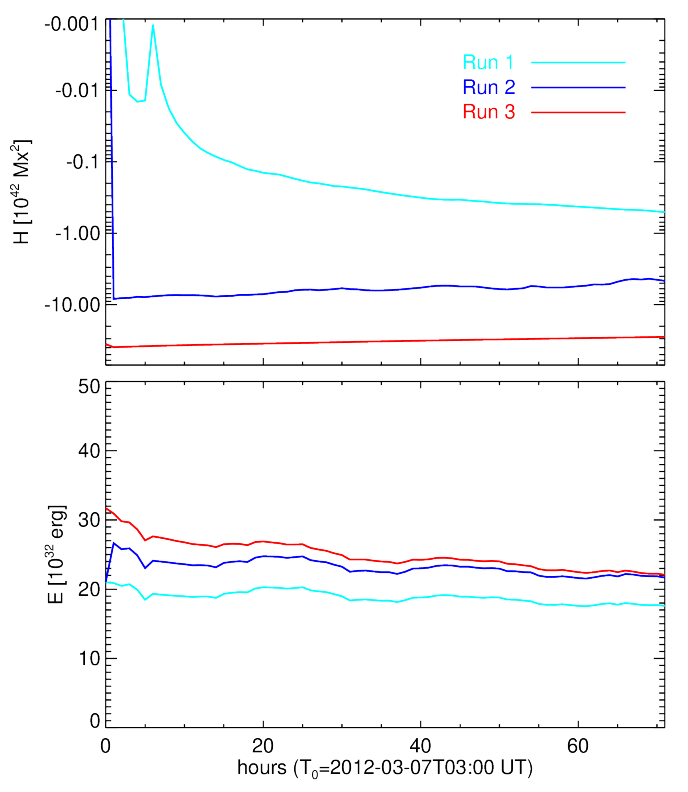}
\caption{Time evolution of top: total magnetic helicity $H$, bottom: total magnetic energy $E$ for \texttt{Run1} (cyan), \texttt{Run2} (blue) and \texttt{Run3} (red). The start time of the simulation is 03:00 UT on March 7, 2012. }
\label{Fig3}
\end{figure*}

\begin{figure*}
	\centering
	\includegraphics[width=.69\textwidth,clip=]{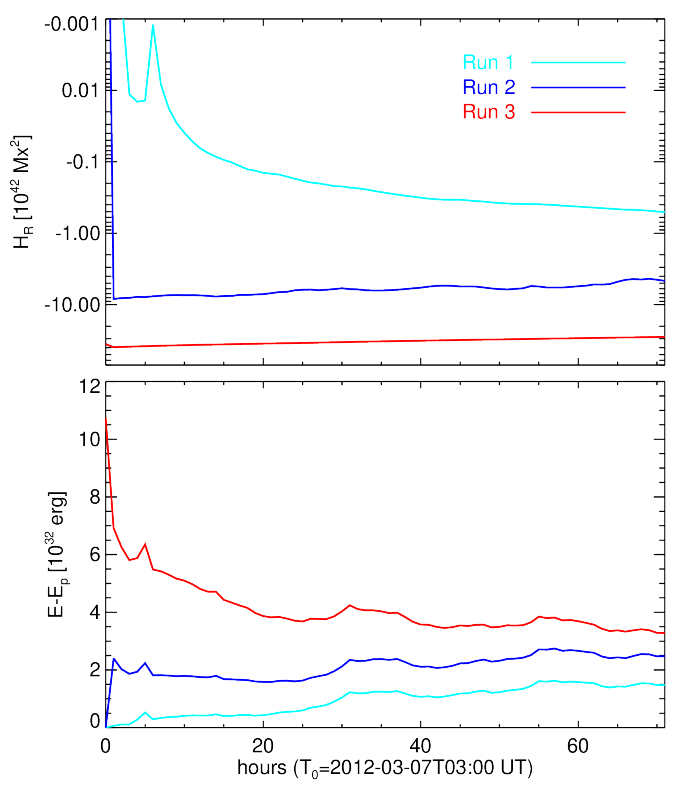}
	\caption{ Time evolution of top: relative magnetic helicity $H_R$, bottom: free magnetic energy $E_f$ for \texttt{Run1} (cyan), \texttt{Run2} (blue) and \texttt{Run3} (red). The start time of the simulation is 03:00 UT on March 7, 2012.  }
	\label{Fig4}
\end{figure*}

\begin{figure*}
	\centering
	\includegraphics[width=.99\textwidth,clip=]{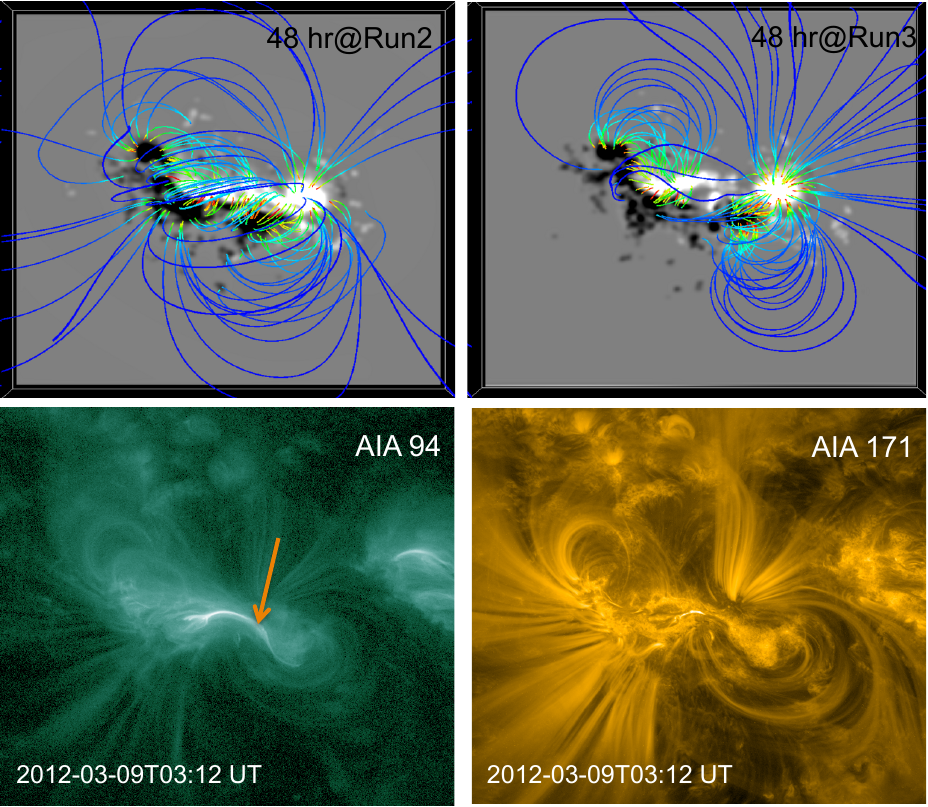}
	\caption{ Comparison of magnetic structure with the coronal plasma tracers in AIA images of the AR 11429. {\bf Top row:} Rendering of magnetic structure at 46th hour of Run2 and Run3. {\bf bottom row:} Images in 94 \AA~show predominant emission along the sheared PIL, which is a signature of a twisted flux rope, whereas 171 \AA~images present plasma loops as the lobes of the sigmoid. The magnetic structure in Run3 better resembles the sigmoid as seen in AIA 171~\AA, however the core field is not as much twisted flux to mimic the hot emission as seen in AIA 94~\AA. }
	\label{Fig5}
\end{figure*} 
\section{Results}
\label{Res}
The AR 11429 was a pre-emerged one and appeared in the north latitudinal belt N17$^o$ as seen from the earth view. Snapshots of vector field observations of the AR 11429 at two different epochs of the evolution are displayed in Fig.~\ref{Fig1}. These magnetograms reveal the presence of a large interface of opposite polarities referred to as the polarity inversion line (PIL). Over time, these polarities evolve with persistent shearing and converging motions. These motions might have led to continuous flux cancellations inferred from the time profile of the magnetic flux in the AR \citep{DhakalS2020}. The coronal EUV images of the AR present a plasma loop structure resembling inverse-S sigmoidal morphology, which was modeled to show the presence of a twisted flux rope in the core of the AR \citep{Vemareddy2021_MagStru}. During the disk passage, the AR produced three major eruptive flares with CMEs of speeds exceeding 1000 km/s. In this study, we consider 72 hours of AR evolution starting from 03:00 UT on March 7, 2012 by driving the initial condition with the time-series of observational data via MF method. After two days of evolution, the AR again produced an eruptive flare of M-class at 03:53 UT on March 9, 2012. This means that the magnetic evolution leads to storage of magnetic energy by building a twisted flux rope, and we aim to capture a similar evolution through MF simulations.

We perform three different MF runs with appropriate combinations of initial and boundary conditions. \textit{Run1} is a simulation with the PF as initial condition driven by \texttt{bc1} boundary condition as described in Sect.~\ref{bc}. \textit{Run2} is the PF initial condition driven by \texttt{bc2} boundary condition. A uniform average twist is derived from the vector field observations and is injected into the corona. In \textit{Run3}, we drive the NLFFF as the initial condition with \texttt{bc3} boundary condition. In this case, an ad-hoc assumption involved is that the magnetic fields decreases exponentially up in the corona, so the horizontal fields are reduced by a $c=0.7$ factor and then used to prescribe the bottom boundary which is changing in time. The 3D magnetic fields of these runs are analyzed and the results are described in the following.

The simulated 3D magnetic field of the runs is visualized with VAPOR software \citep{LiShaomeng2019_vapor}. In Fig.~\ref{Fig2}, the magnetic structure of the AR is displayed by tracing the field lines at different epochs of the evolution. The foot point locations of the depicted field lines are chosen with a bias based on a variable such as horizontal magnetic field ($B_h$) and total current magnitude $J_{tot}$. The initial field structure is shown in the first row panels. Even if the shearing motions of the opposite polarities, the magnetic structure in \texttt{Run1} exhibits less non-potential without much difference from the initial field. We expect that this is due to no magnetic twist information being pumped from the boundary which even with horizontal motions deforms the initial potential field state insignificantly. The magnetic structure in \texttt{Run2} evolves to a little more sheared state since we assume the boundary field to be linear force free. Being already in a non-potential state, the magnetic structure in \texttt{Run3} appears highly sheared and resembles inverse-S shaped as a whole. It can be noticed that the shearedness increases with time because the twist information from the horizontal field is added to driver field from boundary observations. An animation of the three runs together is available online for a better comparison of simulations.

In addition, we have computed magnetic energy and helicity from the volumetric distribution of magnetic field above the AR and plotted in Figure~\ref{Fig3} and~\ref{Fig4} with time. Especially, to compare the coronal helicity budget, we derive the relative magnetic helicity \citep{Berger1984} from the 3D simulated field as given by 
\begin{equation}
	H_{\rm R}=H-H_{\rm p}=\int \mathbf{A}\cdot \mathbf{B}\,dV - \int \mathbf{A_{\rm p}}\cdot \mathbf{B_{\rm p}}\,dV
\end{equation}
Here the reference field is potential field $\mathbf{B}_{\rm p}$ and $\mathbf{A}_{\rm p}$ is the corresponding vector potential, which have same normal component of boundary field as that of the actual magnetic field under evaluation. The magnetic helicity $H$ in \texttt{Run1} increases in magnitude but smaller by two orders (10$^{40}$ Mx$^2$) to that obtained in \texttt{Run2} and \texttt{Run3}, where the $H$ value decreases over the course of the AR evolution by about 25\% from the initial value. The $H$ value varies from $39\times10^{42}$ Mx$^2$ in \texttt{Run3} which is a typical value in orders of magnitude to produce a large-scale CME eruption. The relative helicity $H_{\rm R}$ differs insignificantly with $H$ since $H_p$ is negligibly small and therefore it is not required to calculate $H_{\rm R}$ separately when using \PC for such simulations. 

We note that the total magnetic flux as well as the average twist ($\alpha_{av}$) in the AR decreases in time \citep{DhakalS2020}. The $H$ increases due to shear motions alone which is below 10$^{42}$ Mx$^2$ in \texttt{Run1}, however, there is no additional increasing twist information in the horizontal field observations, and as a result the coronal helicity budget decreases in \texttt{Run2} and \texttt{Run3} from their initial values. Under these conditions, it is difficult to model the formation of twisted flux and its eruption. These type of ARs have to be treated with a provision to pump as much helicity through boundary observations into the corona \citep{Cheung2012_MF}.

We have also evaluated the total and free magnetic energies. Given 3D magnetic field in the computational volume, the free magnetic energy is calculated by
\begin{equation}
	E_{\rm f}=E-E_{\rm p}=\frac{1}{8\pi}\int \left( \mathbf{B}^2-\mathbf{B}_{\rm p}^2 \right)\,dV   
\end{equation}
where $B_{\rm p}$ is the potential magnetic field. The total energy $E$ decreases from $31\times10^{32}\,ergs$  at the start of the \texttt{Run3} to $22\times10^{32}\,ergs$, which is following the total magnetic flux evolution as in Figure~\ref{Fig1}c. This corresponds to decrease of $E_f/E$ from 20\% to 15\%. For \texttt{Run2} and \texttt{Run1}, the $E$ follows a similar trend with lower values compared to \texttt{Run3}. However, $E_f/E$ increases to 10\% for \texttt{Run1} and to 15\% for \texttt{Run2} during the modelled evolution. In all these runs, the free energy is typically comparable to M-class flares, and shows the characteristic difference among the three runs performed with different boundary and initial conditions.

Finally, we also compare the magnetic structure qualitatively with the coronal imaging observations in EUV wavelengths. In the top row of Fig.~\ref{Fig5}, we have displayed the rendering of magnetic structure at the time instance of 46th hour from \texttt{Run2} and \texttt{Run3}. The snapshots at this time correspond to the CME eruption at 3:53 UT on March 9, for which the pre-eruptive AIA 94 and 171~\AA~images at 03:12 UT are displayed in the bottom panels of Fig.~\ref{Fig5}. The plasma tracers in AIA 171~\AA~ present two lobes; one of their legs is adjacent to the other on opposite sides of the PIL. This implies a highly sheared magnetic field resembling an inverse-S sigmoidal configuration \citep{Green2002,Vemareddy2018}. This kind of morphology is almost similar to the magnetic structure generated in \texttt{Run3} than in \texttt{Run2}. However, the predominant emission seen in AIA 94~\AA~along the sheared PIL refers to twisted flux rope at the core of the sigmoid, which is not reproduced in the simulation. Previous MF simulations by \citet{Yardley2018_MF_11437} also showed a similar difficulty of generating twisted flux at the core; however, the lobes of the sigmoid and their evolution were reported to be well captured. Altogether, the MF simulation presented here is able to reproduce the dynamical evolution of the AR consisting of a highly sheared magnetic field globally that is able to launch CMEs. 

\section{Summary and Conclusions}
\label{Summ}
We have performed the data-driven MF simulations of the AR 11429. The numerical procedure of the MF simulation is implemented, for the first time, in \PC, which is publicly available to serve a variety of astrophysical problems. In this case, we are required to use only a magnetic module, and the boundary conditions have to be prepared from the observations of the AR under study in the form of vector potential, which then have to be fed to the bottom boundary layer as the simulation advances in time. The twisted magnetic field generated from the MF simulations may also serve as the initial conditions for the full MHD runs including plasma, where one intends to study the thermodynamic evolution of flaring plasma, etc. Compared to \citet{WP2019}, where the helicity is injected directly at the boundary, this approach is much closer to the observed loop structure and hence is more self-consistent with photospheric input data.

We perform three different MF runs with appropriate combinations of the initial and boundary conditions. In this work, we invented a way to invoke magnetic field observations to drive the coronal field. The key point is to work with vector potential $\mathbf{A}$  than magnetic field $\mathbf{B}$ without worrying about divergence condition. However, there is a lot of difficulty in converting $\mathbf{B}$ to $\mathbf{A}$ and this is the first such effort. The simplest case is to drive the potential field with the observed normal component of magnetic field, which cannot lead to build up of a sheared coronal magnetic structure. The latter two combinations of the initial and boundary conditions are meant to inject magnetic twist information obtained from the vector field observations of the magnetic field. We found that the already sheared field, further driven by the sheared magnetic field, will maintain and further build the highly sheared coronal magnetic configuration, as seen in AR 11429. The sigmoidal morphology exhibited by AIA images at the pre-eruptive time is better reproduced by the boundary condition with twist information added to the vector potential and then drive the NLFFF since the AR is pre-existing with a non-potential magnetic field, even after the first eruption on March 7. The quantitative estimates of magnetic helicity and energy clearly show a marked distinction in these three runs, with the higher of these values corresponding to the complex and twisted nature of the magnetic field.

Although the global sheared configuration is reproduced, the core field is not of twisted flux mimicking the hot plasma emission. The twisted flux along the PIL is not generated at the small scales, which is a drawback to deal with MF simulations in this study, as was also the case in the previous such simulations \citep{Yardley2018_MF_11437}. As a note, the AR 11429 is a decaying active region, so the input of magnetic twist from the observations is less significant to reflect in the driver field, which could be a reason for the decreasing free energy and magnetic helicity. Under these conditions, capturing the formation of twisted flux during the AR evolution is a challenge. In such instances, new approaches with certain ad hoc assumptions are required. \citet{Cheung2012_MF} derived the electric fields based on the time sequence vector magnetic field with free parameters that add the twist information proportionately to the observations. Instead of magnetic fields, the electric fields thus derived were used to drive the MF simulations and then study the formation of the twisted flux rope at the core and its further eruptions. Such simulations have recently been employed \citep{Cheung2015_helicaljets, Pomoell2019_EleFld} to study the formation of the twisted flux rope in the course of AR evolution and then its rise motion in few ARs. Therefore, it is worth driving the MF simulations with electric fields, and our future reports will follow such investigations.

%\section*{Data Availability} The data used in this manuscript is obtained from NASA's SDO mission and is publicly available from Joint Science Operations Center (http://jsoc.stanford.edu/).)

\normalem
\begin{acknowledgements}
PV acknowledges the support from DST through Startup Research Grant. SDO is a mission of NASA's Living With a Star Program. Field line rendering is due to VAPOR visualization software (\url{www.ucar.vapor.edu}) The author also thanks Gherardo Valori for his insights on the construction of vector potentials. This project has received funding from the European Research Council (ERC) under the European Union's Horizon 2020 research and innovation program (Project UniSDyn, grant agreement no 818665) (JW). This work was done in collaboration with the COFFIES DRIVE Science Center. We thank an anonymous referee for helpful comments and suggestions.
\end{acknowledgements}
  
\bibliographystyle{raa}
%\bibliography{ref_lib08112022.bib}

\end{document}